\documentclass[a4paper,10pt,titlepage]{article}
\usepackage[dvips]{graphicx}
\usepackage{array}

\title{A simple model of the Auroral Kilometric Radiation visibility}
\author{R. Schreiber\\ 
Nicolaus Copernicus Astronomical Center PAS\\ 
ul.Rabia\'nska 8, PL 87-100 Toru\'n, Poland}
\date{}

\begin{document}
\maketitle

\begin{abstract}
Data collected with Polrad swept frequency analyzer during almost 2.5 years of Interball-2 mission are used to investigate the directivity properties of AKR observed in the Northern Hemisphere. In contrast with the previous, mostly statistical studies, single dynamic spectra have been analyzed. Simple modeling of AKR source visibility from the location of the observer makes possible to find approximate locations of remotely observed radiation sources and to model the overall form of the dynamic spectrum with a single set of geometric parameters valid during up to tens of minutes. Cases of filled and partially filled emission cones are found as well as smaller emission cones embedded in the main cone. Anisotropic AKR beams aligned with subcones as thin as $10^\circ$ and sometimes narrower have been detected. 
\end{abstract}

\section{Introduction}

Auroral Kilometric radiation has been observed since more than 30 years [\textit{Benediktov et al.}, 1965, \textit{Dunckel et al.,}, 1970,  \textit{Gurnett}, 1974]. Although the body of available information concerning AKR is impressive there are still some very simple questions without a satisfactory answer. A set of basic facts, some established quite a long time ago, will serve as a basis for the present paper.  

\begin{enumerate}
	\item AKR sources are located on magnetic field lines having their footprints within the auroral oval [\textit{Gallagher and Gurnett}, 1979, \textit{Benson and Calvert}, 1979, \textit{Huff et al.}, 1988 and recently \textit{Gurnett et al.}, 2001, \textit{Schreiber et al.}, 2002, \textit{Mutel et al.}, 2003, \textit{Mutel et al.}, 2004] 
	\item Radiation is observed within delimited region [\textit{Green Gurnett and Shawhan}, 1977, \textit{Gallagher and Gurnett}, 1979, \textit{Hashimoto}, 1984]. R-X mode as the main product of electron maser cyclotron mechanism commonly accepted for AKR undergoes strong upward refraction close to the source [\textit{Calvert}, 1981, \textit{de Feraudy and Schreiber}, 1995, \textit{Louarn et al.}, 1996a].
	\item AKR is confined to the seemingly filled [\textit{Green et Gallagher.}, 1985, \textit{Green et al.}, 2004] emission cone, on the other hand cyclotron maser theory predicts strongest amplification of the radiation perpendicularly to the local magnetic field in the source [\textit{Wu and Lee}, 1979, \textit{Pritchett et al.}, 2002] what can suggest a hollow emission cone.
	\item AKR dynamic spectra contain a lot of features, some are rather long-lived (tens of minutes) and aligned with the lower boundary of the dynamic spectrum
\end{enumerate}

\subsection{Problems to be addressed}
Present paper focuses on four primary questions:
\begin{enumerate}
	\item What part of AKR radiating auroral oval can be seen from the single spacecraft location?
	\item How to determine approximate AKR source region location using the shape of dynamic spectrum?
	\item Is the AKR emission cone filled, hollow (or both)?
	\item What can be inferred on the AKR beaming from the dynamic spectrum?
\end{enumerate}
Answers to all above questions rely on simple geometrical reasoning and as it will be shown later, the lack of detailed information about the remote AKR sources and their exact positions will not affect the general conclusions of the paper. 

\section{Instrumentation and data}
AKR dynamic spectra analyzed in the paper belong to the data set collected during almost 2.5 years of Interball-2 mission effective life. Satellite was launched on August 29, 1996 into an elliptical orbit with an inclination of $62.8^\circ$, an apogee altitude of 19140 km, and a perigee altitude of 772 km. Dynamic spectra were recorded by step frequency analyzer Polrad [\textit{Hanasz et al.}, 1998] swept over one of two frequency ranges: 4 kHz - 0.5 MHz, or 4 kHz - 1 MHz, with repetition periods of 6 or 12 s and a frequency resolution of 4 kHz. Three independent channels allowed reception of signals from one or three perpendicular electric antennas ($X$ - monopole 11 m long, $Y$ and $Z$ - dipoles 22 m long). Antenna system was spinning around the axis parallel to the X antenna with a period of 120 s. Polrad was working in one of its operational modes as a polarimeter. One frequency step corresponds to 25 or 50 ms collecting time with 6 ms output filter time constant. The orbit was chosen in such a way that the satellite path in its apogee part was roughly tangent to the auroral oval. Data were recorded during northern passes when spacecraft was above Earth's radiation belts.

\section{Choice of events}
Dynamic spectra with contiguous emissions lasting for tens of minutes were selected. Interball-2 was not crossing the AKR source regions but spectra recorded during source approach and subsequent receding are readily available. Most interesting are spectra with aligned bottom and upper frequency edges. It will be shown that they contain information about filling of the emission cone. On the other hand, bright features lasting for some time (at least a few minutes) can carry information about AKR beaming. Necessary care was taken to avoid proximity of plasmasphere. Refraction and/or reflection caused by plasmasphere obscuring AKR view has not been taken into account in proposed model. 

\section{Model}

Two scenarios of AKR source observations are employed (in both cases R-X mode has been assumed as a basic carrier of AKR energy).  
\begin{enumerate}
	\item \textit{For calculating AKR visibility from a given point of the Interball-2 orbit} a set of radiating magnetic field lines anchored to the auroral oval defined by its invariant latitude is used. Field lines fill the whole 24 hours MLT range. A rectilinear AKR propagation between source and observer is assumed.
	\item \textit{For finding AKR source position on the auroral oval and for emission cone investigation} the source is located on a single magnetic field line. Both the rectilinear propagation to the observer and a simple raytracing (based on \textit{Calvert's}, [1981]) formulas have been taken into account. 
\end{enumerate}
Earth's magnetic field is approximated by a centered dipole. Positions of the AKR sources on individual magnetic field lines are defined assuming escape of EM radiation slightly above the local R-X mode cutoff frequency $f_x$ [\textit{Wu and Lee}, 1979, \textit{de Feraudy and Schreiber}, 1995, \textit{Louarn et Le Qu\'eau}, 1996a, \textit{Ergun et al.}, 2000]. The ratio of AKR escape frequency $f$ to $f_x$ is a free parameter of the model. Second parameter relates to the constant $N_0$ in \textit{Nsumei et al.}, [2003] polar electron density model adopted for this paper. Maximum AKR emission frequency range 0.025 - 0.800 MHz has been adopted, practically it was limited to the observed AKR bandwidth. Frequency range was divided into 250 frequency steps, corresponding points on magnetic field lines were treated as nodes with attached AKR emission cones. Cone axis was tangent to the magnetic field line. 

\subsection{AKR source visibility from a given satellite position}
Simple ``visibility'' diagram for the AKR is proposed. Assuming that the AKR can be produced within the whole MLT range of the auroral oval (in this simplified model labeled by a single invariant latitude value) one can construct for a given position of the observer a contour plot of the propagation angle $\alpha$ (between direction to the observer and the direction antiparallel to the local magnetic field in the source $\bf B$) as a function of emission frequency (or vertical position of the source on the field line) and MLT of that line (Fig.1). Having in mind that R-X mode in the source region is generated most efficiently at angles $\approx 90^\circ$ to $\bf B$ [\textit{Hilgers, de Feraudy and Le Qu\'eau}, 1992, \textit{Ergun et al.}, 1998, \textit{Ergun et al.}, 2000], and refracted upward in the vicinity of the source, radiation cannot move out of a half-sphere. In the real situation $\alpha < 90^\circ$ [\textit{Bahnsen et al.}, 1987, \textit{Roux et al.}, 1993, \textit{Louarn et Le Qu\'eau}, 1996a]. R-X mode must climb up the generation cavity before escaping to the free space.  That leads to upward rotation of $\bf k$ vector. Similar additional effect can be expected as a result of refraction at the cavity edges. Therefore line $\alpha = 90^\circ$ delimits on $f - MLT$ plane region of AKR visibility. 

The trivial property of the diagram is its symmetry with respect to the observers position in MLT. It results from assumption of constant invariant latitude for all parts of the oval. Another property: the lowest observable frequency cannot be lower than the  R-X mode cutoff at the observers location. For remote observations it is always higher than that. Partially filled emission cone, when AKR is radiated for a range of emission angles, will map for fixed source position to the limited frequency range on the diagram. For a given frequency a contiguous part of the oval (or two segments in case of partially filled emission cone) can eventually host the AKR source.

``Visibility'' diagram can be used:
\begin{enumerate}
	\item for single satellite observations to decide what parts of the oval eventually radiating AKR can (or cannot) be seen from the observer's location. Such diagrams can be quite distinct for different satellites: low orbiting FAST will ``see'' at a given position only a few hours segment while POLAR could gather AKR radiation generated probably in many sources over most of the oval
	\item for simultaneous observations from several spacecraft to define intersection of visibility diagrams and locate regions on the oval producing AKR eventually accessible for all spacecraft. Different satellites are usually hit by different beams of AKR and the intersection of their visibility diagrams may correspond to different propagation angles and azimuths
	\item for approximate determination of emission cone opening angles (taking into account the instantaneous AKR frequency range).
\end{enumerate}

Keeping in mind simplicity of the proposed approach such results should be treated as informative but sometimes they can be quite useful, for example for restricting the positions of possible AKR source regions or for taking decision whether remotely observed AKR is produced on dayside or nightside part of the oval. 

\subsection{Modeling of the shape of AKR dynamic spectrum and finding position of the source on auroral oval}
AKR ``visibility'' diagram offers a simple geometrical information concerning the radiation sources when there is no good data about the source position on the oval. Having (or assuming) that position it is possible to reproduce overall shape of AKR dynamic spectrum.  Every modeled emission cone consists of a set of embedded subcones parametrized by their propagation angle $\alpha$. For upgoing rays $\alpha$ is smaller than $90^\circ$. The basic geometric outline is demonstrated in Fig.2.  For the raytracing purpose angle $\theta$ (between $\bf k$ vector and the direction antiparallel to the local magnetic field in the source $\bf B$) analogical to $\alpha$ will be introduced.  

For equally spaced points on the orbit (in this paper one minute apart) and frequencies (250 steps for dynamic spectrum frequency range) one has to calculate for the rays reaching the satellite the values of $\alpha$ and $\theta$ angles at the exit of the radiation from AKR source. Angle $\alpha$ is calculated from the dot product of the $\bf B$ and direction to the observer. More complicated task of finding $\theta$ angles is based on the interpolation between many rays launched from the source. Curves of constant $\alpha$ (and $\theta$) angle will be overlaid on the dynamic spectrum. Fig.3 shows outline of projection of two $\alpha = const$ curves onto the dynamical spectrum. In the present paper analytic raytracing has been based on formulas published by \textit{Calvert} [1981]. His dispersion relation for the extraordinary waves in a low-density plasma has the form:
  
\begin{equation} \label{R}
{{}{ n}^{ 2}_{ z} + \left({ R + 1}\right) \cdot {}{ n}^{ 2}_{ x} / 2 = R}
\end{equation} 

where

\begin{displaymath}
{ R = 1 - X / \left({ 1 - Y}\right)}
\end{displaymath}
\begin{displaymath}
{ X = {}{ \left({ { f}_{ pe} / f}\right)}^{ 2}}
\end{displaymath}
\begin{displaymath}
{ Y = { f}_{ ce} / f}
\end{displaymath} 

and ${ { f}_{ pe}}$ is the plasma frequency, ${ { f}_{ ce} }$ equals electron gyrofrequency and ${f}$ is the AKR frequency.
 
Refraction index components ${n_x}$ and ${n_z}$ are correspondingly perpendicular and parallel to the magnetic field. For rays launched perpendicularly to the magnetic field ${n_z} = 0$ at the beginning of ray path. Equation(\ref{R}) defines ${R_0}$ value at that level:

\begin{equation} \label{R0}
{ { R}_{ 0} = \left({ { R}_{ 0} + 1}\right) \cdot \frac{ {}{ n}^{ 2}_{ x}}{ 2}}
\end{equation}

More general form of equation (\ref{R}) should explicitly take into account oblique propagation with nonzero ${n_x}$ and ${n_z}$ starting components. 

\begin{equation} \label{R00}
{ { R}_{ 0} = \left({ { R}_{ 0} + 1}\right) \cdot \frac{ {}{ n}^{ 2}_{ x}}{ 2} + {}{ \left({ \frac{ { n}_{ x}}{\tan   \theta}}\right)}^{ 2}}
\end{equation}

Solving (\ref{R00}) with respect to ${n_x}$ and substituting the result to (\ref{R}) gives:

\begin{equation} \label{nznx}
{ \frac{ { n}_{ z}}{ { n}_{ x}} = \sqrt{{ \frac{ R \cdot \left({ 1 + 2 \cdot {}{ \left({ {\tan \theta}}\right)}^{ - 2}}\right) - { R}_{ 0}}{2 \cdot { R}_{ 0}}}}}
\end{equation}

As in Calvert's paper raytracing is based on the Snell's law and the differential equation for ray paths will result from the equality  ${dz / dx = n_z / n_x}$. There are two consequences of this approach:

\begin{enumerate}
\item no lateral deviations of the rays are taken into account
\item wave normal and group velocity vectors are supposed to be parallel, this is certainly true close to the directions parallel or perpendicular to the magnetic field, for the intermediate angles and the set of plasma parameters adopted in the paper the angles difference can attain up to a few degrees. As in Calvert's paper comparisons were made with results of 3-D raytracing code based on Haselgrove set of differential equations [\textit{de Feraudy and Schreiber}, 1995], and no significant discrepancies were found.
\end{enumerate}  

Using Calvert's substitutions for ${X}$ and ${Y}$ in ${R}$ definition ($X = const$, $Y = 1 - A \cdot d$ and ${ z = A \cdot d / X}$ where $A$ is magnetic field gradient and $d$ distance along field line) equation (\ref{nznx}) can be reformulated to its final form:

\begin{equation}
{ \frac{ dx}{ dz} = \sqrt{{ \frac{ 2 \cdot z \cdot \left({ { z}_{ 0} - 1}\right)}{ z + 2 \cdot { z}_{ 0} \cdot {}{\left({ {\tan \theta}}\right)}^{ - 2} \cdot \left({ z - 1}\right) - { z}_{ 0}}}}}
\end{equation}

where $z_0 = 1 / (1 - R_0)$ 

This differential equation has a rather cumbersome closed form integral subsequently used for ray paths calculation. 

Rays span the full $360^\circ$ azimuthal angular range around radiating magnetic field line. For every ray arriving at the satellite its azimuth with respect to the magnetic shell surface element has been determined (starting from West through South to East). A set of constant azimuth values curves has been overlaid on the dynamic spectrum, results are shown in Fig.4 and are discussed later. There can be problems with determination of absolute azimuth in case of folded source region (see scenario proposed by \textit{Louarn and Le Qu\'eau}, 1996a, 1996b or \textit{Pritchett et al.}, 2002) but the azimuth values together with the range of $\alpha$ give an idea about extension of the AKR beam producing observed dynamic spectrum. Refraction plays an important role in rays deflection close to the AKR source [\textit{de Feraudy and Schreiber}, 1995] and in general case of the spacecraft not crossing the source region the $\theta$ curves will be placed on the dynamic spectrum above the $\alpha$ set. The $\alpha$ curves can be subsequently used for the estimate of emission cone opening angle. It is possible to look for eventual dependency of emission cone opening angle on the frequency (within the frequency band of the dynamic spectrum). Fits of the curves to the shape of the dynamic spectrum, especially to its lower frequency boundary, depend in this model on four parameters:  $N_0$ and $f / f_x$, the invariant latitude and MLT of the source region. Initial guess of source MLT coordinate relates to satellite MLT for the lowest point of the spectrum (if bottom part of the spectrum has a typical saucer-like form). $N_0$ controls mainly vertical position of the curves and to some extent curvature.  In the present model $N_0$ varies within the limits 500 - 2000 $cm^{-3}$ and  $f/f_x$ between $1.01$ and $1.10$ (based on estimates by \textit{de Feraudy and Schreiber}, 1995 and  \textit{Louarn and Le Qu\'eau}, 1996b). Changes of invariant latitude act in more complicated way: they can alter curvature but also rotate curves and change vertical position of the whole set. Similar transformations can be observed while changing MLT value. For cases presented in this paper good fits can be produced within a box $\pm 0.5 \div 1.0$ hour in MLT (depending on the case) and $\pm 2^\circ$ in invariant latitude. For a fixed source position on the oval variations of $N_0$ and $f/f_x$ change only slightly positions of $\alpha$ curves while the $\theta$ curves can move appreciably within the frame of the dynamic spectrum. It can be expected for the source region height determined mainly by the approximate equality of the AKR frequency and the local electron gyrofrequency $f_{ce}$ in the source region. That means relatively good determination of emission cone opening angle even without good knowledge of the $f/f_x$, and $N_0$ values and of initial $\bf k$ vector direction. On the other hand, the fit of $\theta$ curves to the dynamic spectrum takes into account R-X mode refraction close to the AKR source and gives additional refinement of the deduced source position.

\subsection{AKR emission cone: is it fully or partially filled?}
Interball-2 spacecraft trajectory is often tangential to the auroral oval. Two basic geometrical situations can be expected: 
\begin{enumerate}
\item tangent crossing of the filled emission cone
\item tangent crossing of the empty (or partially filled) emission cone.
\end{enumerate}
Both situations are demonstrated in Fig.3. For filled emission cone the upper frequency (dashed line) does not change much with time (assuming approximately stable position of the AKR source bottom). But if the empty part of emission cone will be traversed, it can be noticed in the overall form of the dynamic spectrum (continuous line representing upper spectrum boundary). The bottom and upper part of the spectrum will be aligned, there will be no constant upper frequency bound. Lines $\alpha = const$ overlaid on the dynamic spectrum can be used to measure actual size of the emission cone. Their intersections with lines $f = const$ gives information about the size of the emission cone produced at a given frequency and traversed by the observer. In most cases satellite will cross only part of the emission cone, if this part is filled, no conclusion concerning the remaining part of the cone can be reached.

\subsection{Structures in the spectrum and the AKR beaming}
There are still two other interesting possibilities:
\begin{enumerate}
	\item narrowband structures inside the dynamic spectrum aligned with $\alpha = const$ or $\theta = const$ lines
	\item narrowband structures at the bottom of the dynamic spectrum
\end{enumerate}
First case corresponds to the AKR emissions beamed into cone with narrow edges, extension in time can correspond to the natural duration of the process generating the structure or to the azimuthal extension of the beam. Structures at the bottom of the dynamic spectrum can represent caustics produced for the rays emitted close to $\theta = 90^\circ$ [\textit{de Feraudy and Schreiber}, 1995].

\section{Results}
Some representative cases illustrate in Fig.4 the classification based on the model.

{\textbf{(a)  24/07/98 0600-0700 UT}}.
 This is a ``typical'' Polrad's spectrum with rather wide emission cone, probably partially empty and attaining $\sim 35^\circ$ wide filling. Bright (more than 20-30 dB above surrounding background) wideband emissions lasting for about 20 minutes and corresponding to a beam about $10^\circ-20^\circ$ wide in opening angle are also visible.
 
{\textbf{(b)  15/08/97 0600-0700 UT}}
 - about $40^\circ-50^\circ$ wide very strong structure at the beginning of dynamic spectrum. Azimuthal range is not well defined and slight changes $\pm0.5$ hour in MLT can move the azimuthal lines pattern up to few tens of degrees since satellite is moving close to the AKR source region. The low frequency part resembles broadband electrostatic emissions, but most probable explanation is the nonlinear response of the receiver to the very strong signal. This interpretation is supported by weak harmonic structures visible above the most intense part of the dynamic spectrum. 

{\textbf{(c)  29/03/98 1700-1800 UT}}
 - quite extended emission angles range (about $50^\circ$ with empty cone between $0^\circ$ and $\approx 20^\circ$), but with a $20^\circ$ azimuthal gap around $70^\circ$. It is worthwhile to note nice alignment of constant azimuth lines with vertical borders of the dynamic spectrum. The stepwise rise of the dynamic spectrum lowest  frequency ($\sim 43 min$) illustrates the sudden change of the R-X mode emission cone opening angle, the much weaker background radiation below propagates in L-O mode (unpublished results of Polrad polarimeter provided by M. Panchenko)  

{\textbf{(d)  20/07/97 1300-1400 UT}}
 - narrow empty emission cone about $15^\circ$ thick, bright structure inside less than  $10^\circ$ wide, up to four fine structures as thin as $1^\circ-2^\circ$, quite extended (about $60^\circ$) azimuth range. Dayside source.

{\textbf{(e)  30/10/97 0000-0100 UT}}
- empty emission cone about $10^\circ$ thick embedded inside diffuse AKR radiation.

{\textbf{(f)  27/05/97 1600-1700 UT}}
- two separate AKR emissions containing structures aligned with $\theta = const$ lines about $2^\circ$ and $10^\circ$ wide.

Visibility diagrams for all above mentioned cases at the satellite closest approach to the source are shown in Fig.5. The shaded area corresponds to the frequency ranges for which AKR is not seen in the spectrum. Table 1 presents best fit parameters.

\section{Discussion}

Averaged intensity variations across AKR emission cone were extensively studied by \textit{Green et Gallagher}, 1985. Conclusions based on such statistical approach generally favor filled emission cone although with different opening angles varying with frequency. Latest findings by \textit{Green et al.}, 2004 show seasonal and solar cycle variations of the AKR dynamic spectra additionally complicating interpretation of emission cone geometry. In such a situation statistical approach wipes out eventual individual properties of the AKR emissions. There were attempts to interpret some peculiarities of \textit{Green and Gallagher}, 1985 results, especially for lower frequencies (56.2 kHz and 100 kHz), as pointing toward empty emission cone [\textit{Calvert}, 1987 or two-beam emission pattern \textit{Zarka}, 1988]. 

The overall form of all dynamic spectra presented in the paper can be delineated by a single set of $\alpha$ lines. Interball-2 missed AKR source regions and calculated positions of the sources are approximate. Positions confined to an error box $\pm 1.0$ hour in MLT and $\pm 2^\circ$ in invariant latitude implicate relatively small source regions stable during at most tens of minutes although sometimes different frequency bandwidths can be excited. They agree with the present estimates made by \textit{Mutel et al.}, 2004 for the short AKR bursts. They report position of AKR source centroid wandering during 15 minutes within a region constrained by 2100 MLT $\pm 1$ h and $70^\circ \pm 3^\circ$ invariant  latitude. The curves $\alpha = const$ fitting the bottom limits of the dynamic spectra give cone half-opening angles for the dynamic spectrum frequency band. Highest $\alpha$ value reported by \textit{Louarn and Le Qu\'eau}, 1996a attained about $70^\circ$.  For cases presented in this paper they vary between almost $90^\circ$ (when the satellite is located close to the source region) and $60^\circ$. The values close to $90^\circ$ may be misleading. The sets of curves overlaid on the dynamic spectra are calculated for the best fit of filamentary source aligned with the specified magnetic field line. If the sources are scattered within the error box then real propagation angles can be slightly different because close to the source region refraction may still bend rays [\textit{de Feraudy and Schreiber}, 1995] and the estimated propagation angles will differ from those measured at greater distances. For remote sources the lines corresponding to $90^\circ$ and $80^\circ$ are always much closer to each other, than the lines representing smaller angles. For the analysis of structures within the dynamic spectrum an error in determination of the cone opening angle does not matter much, structure $10^\circ$ wide can shrink to say $7.5^\circ$ or expand to $15^\circ$ depending on their vertical positions on the dynamic spectrum but this is still a relatively narrow beam. For AKR sources close to the satellite as in Fig.4b and Fig.4c both sets of lines do not differ much. 

Good alignment of $\alpha = const$ lines with bright structures in dynamic spectrum implies constant opening angle of the emission cone (or embedded subcone)  within the frequency band occupied by the structure.  

Azimuthal extension of the radiation beam corresponding to the bright structures visible in analysed AKR spectra vary from a few degrees to more than $80^\circ$. At the present stage it is not possible to separate space and time effects and independent observations from other spacecraft located in that same beam would be helpful. But at least the minimum azimuthal extension of the beam can be estimated. It is difficult to compare our beam sizes with AKR short bursts beams discussed recently by \textit{Mutel et al.}, 2004 because our findings suggest rather small $\sim 10^\circ$ latitudinal thickness of the bright parts of emission cone what suggests anisotropic beams. \textit{Mutel et al.}, 2004 comments on AKR burst beams sizes in the frame of Cluster spacecraft separation as rarely ($10\div30\%$) reaching $20^\circ$ but apparently does not take into account possible selection effects introduced by orientation of the baseline projection on the emission cone edge. 

Some fine structures ($\sim 1^\circ - 2^\circ$ wide) observed by Polrad can be interpreted as produced by thin conical beams located on the wall of the hollow emission cone. They resemble thin Jovian DAM arcs recently discussed by \textit{Kaiser et al.}, 2000. 

\section{Conclusions}
Dynamic spectra investigated on case by case basis contain a lot of information concerning radiation diagram of AKR complementary to information gathered in previous statistical studies. Simple model presented above is sufficiently good to catch and reproduce important characteristics concerning AKR directivity. 

\begin{enumerate}
	\item General shape of the AKR dynamic spectrum can be modeled during tens of minutes with one set of model parameters what implies stability of the overall source size and position within the $\pm 1$ h and $\pm 3^\circ$ error box during that period of time. It is possible to determine approximate position of the AKR source region even for remote observations.
	\item In many situations AKR emission cone is not filled, judging from the cases discussed in the paper, ``walls'' can be $10\div 60^\circ$ thick.
	\item Some strong (20 db and more above the AKR background level) and relatively narrowband (a few tens of kHz wide) structures lasting for about ten minutes correspond to thin parts of emission cone ($\approx 10^\circ$ wide). Other structures a few kHz wide can be produced by very narrow beams attaining $1^\circ - 2^\circ$  like those reported for Jupiter DAM arcs.
	\item Structures in the dynamic spectra aligned with constant emission cone opening angles are consistent with the presence of anisotropic AKR beams (subcones) restricted both in elevation and in azimuth.   
	\item Alignment of the curves of constant propagation ($\alpha$) and emission ($\theta$) angle with low frequency borders of the dynamics spectrum or with bright structures inside the spectrum implies constant opening angle of the emission cone for the range of frequencies covered or corresponding altitude range of the source regions.
	\item There are dynamic spectra showing gaps in time what can eventually be interpreted as a result of emission cone ``bitten out'' in azimuth. Confirmation of this conjecture needs however multispacecraft observations.  
\end{enumerate} 

{\bf Acknowledgments}\\
This paper is a result of a rather lengthy process and I am very much indebted to many people I was discussing the problem with, in particular to Alain Hilgers, Herve de Feraudy, Jan Hanasz, Wynne Calvert, and Helmut Rucker. This work was supported by the Committee of the Scientific Research in Poland, grant no. 5 T12E 001 22.

\newpage
\begin{figure}
\caption{Example of AKR visibility diagram for Polrad on 24/07/98  0630 UT. Hatched regions correspond to the parts of the auroral oval not visible from the momentary observer location}
\end{figure}

\begin{figure}
\caption{AKR emission cone/embedded subcone concept}
\end{figure}

\begin{figure}
\caption{AKR emission cone/subcones and beams aligned with cone edges projected on the dynamic spectrum.}
\end{figure}
 
\begin{figure}
\caption{AKR dynamic spectra with superposed $\theta = const$ - solid line, $\alpha = const$ - dashed line and azimuth angle - dotted line}
\end{figure}

\begin{figure}
\caption{Visibility diagrams for the spectra shown in Fig.4 at the satellite closest approach to the source.  The shaded areas correspond to the frequencies not seen in the spectrum, hatched regions cannot be seen at the given observer location}
\end{figure}

\begin{table}[!hbp]
\begin{center}
\begin{tabular}[c]{|c|c|r|r|c|} \hline
Date     & InvLat &  MLT   & $f / f_x$ & $N_0$           \\ \hline
24/07/98 & $69^\circ$         & 19.3 h &  1.05     & $2000 \;cm^{-3}$  \\ \hline
15/08/97 & $70^\circ$         & 20.5 h &  1.10     &  $500 \;cm^{-3}$  \\ \hline
29/03/98 & $67^\circ$         &  2.8 h &  1.10     &  $500 \;cm^{-3}$  \\ \hline
20/07/97 & $68^\circ$         &  8.0 h &  1.05     & $2000 \;cm^{-3}$  \\ \hline
30/10/97 & $70^\circ$         & 17.5 h &  1.05     &  $500 \;cm^{-3}$  \\ \hline
27/05/97 & $70^\circ$         &  3.0 h &  1.10     &  $500 \;cm^{-3}$  \\ \hline
\end{tabular}
\end{center}
\caption{\bf Source parameters}
\end{table}


\begin{thebibliography}{}

\bibitem{} Bahnsen, A. B., M. Jespersen, E. Ungstrup, and I. B. Iversen (1987), Auroral hiss and kilometric radiation measured from the Viking satellite, \textit{Geophys. Res. Lett., 14}, 471.

\bibitem{} Benediktov, E.A., G.G. Getmantsev, Yu.A. Sazonov and A.F. Tarasov (1965), Preliminary results of measurement of the intensity of distributed extraterrestrial radio-frequency emission at 72 and 1525 kC frequencies by the satellite Electron 2, \textit{Cosmic Research, 36, (6)}, 791.  

\bibitem{} Benson, R. F., and W. Calvert (1979), ISIS 1 observations at the source of Auroral Kilometric Radiation, \textit{Geophys. Res. Lett., 6}, 479.

\bibitem{} Calvert, W. (1981), The signature of Auroral Kilometric Radiation on Isis 1 Ionograms, \textit{J. Geophys. Res., 86}, 76.

\bibitem{} Calvert, W. (1987), Hollowness of the observed auroral kilometric radiation pattern, \textit{J. Geophys. Res., 92}, 1267.

\bibitem{} Dunckel, N., B. Ficklin, L. Borden, and R. A. Helliwell (1970), Low-frequency noise observed in the distant magnetosphere with OGO-1, \textit{J. Geophys. Res., 75}, 1854.

\bibitem{} Ergun, R. E., C. W. Carlson, J. P. McFadden, F. S. Mozer, G. T. Delory, W. Peria, C. C. Chaston, M. Temerin, R. Elphic, R. Strangeway, R. Pfaff, C. A. Cattell, D. Klumpar, E. Shelley, W. Peterson, E. Moebius, and L. Kistler (1998), FAST satellite wave observations in the AKR source region, \textit{Geophys. Res. Lett., 25}, 2061. 

\bibitem{} Ergun, R. E., C.W. Carlson, J. P. McFadden, G. T. Delory, R. J. Strangeway, and P. L. Pritchett (2000), Electron-cyclotron maser driven by charged-particle acceleration from magnetic field-aligned electric fields, \textit{Astrophys. J., 538}, 456.

\bibitem{} de Feraudy, H., R. Schreiber (1995), Auroral radiation ray distribution in the light of Viking observations of AKR, \textit{Geophys. Res. Let., 22}, 2973.

\bibitem{}  Gallagher, D. L., and D. A. Gurnett (1979), Auroral kilometric radiation: Time-averaged source location, \textit{J. Geophys. Res., 84}, 6501. 

\bibitem{} Green, J. L., D. A. Gurnett, and S. D. Shawhan (1977), The angular distribution of auroral kilometric radiation, \textit{J. Geophys. Res., 82}, 1825.

\bibitem{} Green, J. L., and D. L. Gallagher (1985), The detailed intensity distribution of the AKR emission cone, \textit{J. Geophys. Res., 90}, 9641.

\bibitem{} Green, J. L., S. Boardsen, L. Garcia, S. F. Fung, and B. W. Reinisch (2004), Seasonal and solar cycle dynamics of the auroral kilometric radiation source region, \textit{J. Geophys. Res. 109}, A05223, doi:10.1029/2003JA010311.

\bibitem{} Gurnett, D. A. (1974),  The Earth as a radio source: Terrestrial kilometric radiation,\textit{ J. Geophys. Res., 79}, 4227.

\bibitem{} Gurnett D. A., R. L. Huff, J. S. Pickett, A. M. Persoon, R. L. Mutel, I. W. Christopher, C. A. Kletzing, U. S. Inan, W. L. Martin, J.-L. Bougeret, H. St. C. Alleyne, and K. H. Yearby (2001), First results from the Cluster wideband plasma wave investigation, \textit{Annales Geophysicae, 19}, 1259.

\bibitem{} Hanasz, J., R. Schreiber, H. de Feraudy, M.M. Mogilevsky, and T.V. Romantsova (1998), Observations of the upper frequency cutoffs of the Auroral Kilometric Radiation, \textit{Annales Geophysicae, 16}, 1097.

\bibitem{} Hilgers, A., H. de Feraudy, and D. Le Qu\'eau (1992), Measurement of the direction of the auroral kilometric radiation electric field inside the sources with the Viking satellite, \textit{J. Geophys. Res., 97}, 8381.

\bibitem{} Huff, R.L., W. Calvert, J.D. Craven, L.A. Frank, and D.A. Gurnett (1988), Mapping of auroral kilometric radiation sources to the aurora, \textit{J. Geophys. Res., 93}, 11, 445.

\bibitem{} Kaiser, M. L., P. Zarka, W. S. Kurth, G. B. Hospodarsky, and D. A. Gurnett (2000), Cassini and Wind stereoscopic observations of Jovian nonthermal radio emissions: Measurement of beam widths, \textit{J. Geophys. Res., 105}, 16053.

\bibitem{} Louarn, P. and D. Le Qu\'eau (1996a), Generation of the Auroral Kilometric Radiation in plasma cavities - I. Experimental study, \textit{Planet. Space Sci., 44}, 199.

\bibitem{} Louarn, P. and D. Le Qu\'eau (1996b), Generation of the Auroral Kilometric Radiation in plasma cavities - II. The cyclotron maser instability in small size sources, \textit{Planet. Space Sci., 44}, 211.

\bibitem{} Mutel, R. L., D. A. Gurnett, I. W. Christopher, J. S. Pickett, and M. Schlax (2003), Locations of auroral kilometric radiation bursts inferred from multispacecraft wideband Cluster VLBI observations. 1: Description of technique and initial results, \textit{J. Geophys. Res., 108}(A11), 1398, doi:10.1029/2003JA010011.

\bibitem{} Mutel, R. L., D. A. Gurnett, and I. W. Christopher (2004), Spatial and temporal properties of AKR burst emission derived from Cluster WBD VLBI studies, \textit{Annales Geophysicae, 22}, 2625.

\bibitem{} Nsumei, P. A., X. Huang, B. W. Reinisch, P. Song, V. M. Vasyliunas, J. L. Green, S. F. Fung, R. F. Benson, and D. L. Gallagher (2003), Electron density distribution over the northern polar region deduced from IMAGE/radio plasma imager sounding, \textit{J. Geophys. Res., 108}(A2), 1078, doi:10.1029/2002JA009616.

\bibitem{} Pritchett, P. L., R. J. Strangeway, R. E. Ergun, and C. W. Carlson (2002), Generation and propagation of cyclotron maser emissions in the finite auroral kilometric radiation source cavity, \textit{J. Geophys. Res., 107}(A12), 1437, doi:10.1029/2002JA009403.

\bibitem{} Roux, A., A. Hilgers, H. de Feraudy, D. Le Qu\'eau, P. Louarn, S. Perraut, A. Bahnsen, M. Jespersen, E. Ungstrup, and M. Andr\'e (1993), Auroral kilometric radiation sources: In situ and remote observations from Viking, \textit{J. Geophys. Res., 98}, 11657.

\bibitem{} Schreiber, R., O. Santolik, M. Parrot, F. Lefeuvre, J. Hanasz, M. Brittnacher, and G. Parks (2002), AKR source characteristics using ray tracing techniques, \textit{J. Geophys. Res., 107}(A11), 1381, doi:10.1029/2001JA009061.

\bibitem{} Wu, C. S. and  L. C. Lee (1979), A theory of the terrestrial kilometric radiation, \textit{Astrophys. J., 230}, 621.

\bibitem{} Zarka, P. (1988), Beaming of planetary radioemissions, in \textit{Planetary Radioemissions II}, edited by H. O. Rucker et al., Austrian Acad. Sci. Press, Vienna, 327.

\end{thebibliography}
\end{document}